\documentclass[twocolumn,showpacs,superscriptaddress,amsmath,amssymb]{revtex4}

\usepackage{graphicx}
\usepackage{color}
\usepackage{dcolumn}
\usepackage{bm}
\usepackage{multirow}

\newcommand{\ignore}[1]{}
\newcommand{\K}{{\mathcal K}}

\begin{document}

\title{Effect of memory in non-Markovian Boolean networks}

\author{H. Ebadi}
\affiliation{Bioinformatics, Institute for Computer Science, Leipzig University,
H\"{a}rtelstrasse 16-18, 04107 Leipzig, Germany
}

\author{M. Saeedian}
\affiliation{Department of Physics, Shahid Beheshti University, G.C., Evin, Tehran 19839, Iran}

\author{M. Ausloos}
\affiliation{School of Management, University of Leicester, University Road,
Leicester LE1 7RH, UK}
\affiliation{eHumanities group,
Royal Netherlands Academy of Arts and Sciences (NKVA),
Joan Muyskenweg 25, 1096 CJ Amsterdam, The Netherlands}
\affiliation{GRAPES, rue de la Belle Jardiniere 483, B-4031, Angleur, Belgium}

\author{G.~R. Jafari}
\affiliation{Department of Physics, Shahid Beheshti University, G.C., Evin, Tehran 19839, Iran}
\affiliation{The Institute for Brain and Cognitive Science (IBCS), Shahid Beheshti University, G.C., Evin, Tehran 19839, Iran}
\affiliation{School of Biological Sciences, Institute for Research in Fundamental Sciences (IPM), Tehran, Iran}

%

\date{\today}

\begin{abstract}

One successful  model of interacting biological systems is the Boolean network.
The dynamics of a Boolean network,  controlled with Boolean functions,  is usually considered to be a Markovian (memory-less) process.
However, both self organizing features of  biological phenomena and their intelligent nature should raise some doubt  about ignoring the history of their time evolution. Here, we extend the Boolean network Markovian approach: we involve the effect of memory on the dynamics. This can be explored by modifying Boolean functions  into non-Markovian functions, for example, by  investigating the usual non-Markovian threshold function, - one of the most applied Boolean functions. By applying the non-Markovian threshold function on the dynamical process of a cell cycle network, we discover a power law memory with a more robust dynamics than the Markovian dynamics.

\end{abstract}


\maketitle

\section{Introduction}

Many interacting systems have been modeled with Boolean networks \cite{Kauffman:2003,Aldana:2003,page1990self,green2007emergence}. This approach  has been  successfully applied to study   biological signaling systems as  corner stones of a wide range of vital phenomena \cite{Alon:2006,Bornholdt:2005,Koch:1983,ebadi:2014,Davidich:2008,Rybarsch:2012}.
The binary values of the network nodes are systematically updated through {\it ad hoc}  Boolean functions. These Boolean functions are linear or non linear combinations of   logical rules. The dynamical path of a Boolean networks is followed  by iterating the updating Boolean functions. In the  biological   models,  the nodes are considered  to be   the molecules;  the biochemical signalings are the (often directed)  links. Specifically,  in gene regulatory networks, the expression level of genes, taken  as the nodes, are discretized as ``all or nothing''. The interactions are classified to be  either positive or negative,  corresponding to activating   or  inhibitory relations between nodes,  respectively. Such a discretization approach has the benefits  to reduce a complicated interacting system to a simple dynamical binary graph, i.e. a    ``Boolean network''. The dynamical features of the Boolean networks have been  widely studied from various points of view: mainly, their path way,  their final state(s),  and their stability \cite{klemm2005,ebadi:2014,Boldhaus:2010,drossel-review}.

Despite the capability of Boolean networks to model the time evolution of the gene regulatory networks, there exists an important ignored fact, thereby previously neglected;  the time evolution of biological systems is affected by the history of their dynamics. Indeed, it should seem  obvious that  the biological process of systems which are claimed to be intelligently designed \cite{dembski2008,meyer2009signature}, should memorize some information about the history of their time evolution. Therefore, the present status of a biological system should not be only dependent of its immediate prior state, -thus memory-less, but  should also  depend  of its further former states, - thus with inclusive memory. Boolean functions, however, up to now, appear to  govern the dynamics of the Boolean network only based on the last state of the network.

Here, we discuss the effect of memory in the present and future states of Boolean networks. In order to examine our hypothesis, we present a case study on YCC, - a problem  which has been widely simulated with Boolean networks \cite{Lau:2007,Li:2004,Davidich:2008,Rybarsch:2012,ebadi:2014}. We specifically focus on the most used class of Boolean functions,  namely \textit{Threshold functions},  by applying some modifications (in Section III) in order to  construct  a non-Markovian threshold function. Finally, we perform a comparative study on the dynamical behavior  resulting from either  two threshold functions. Interestingly, we  reach  a more robustness output than the one obtained  for non-Markovian threshold functions. The effect of memory appears to follow a power law which guarantees the scalability of the process in  time.

\begin{figure} 
\includegraphics[scale=0.5]{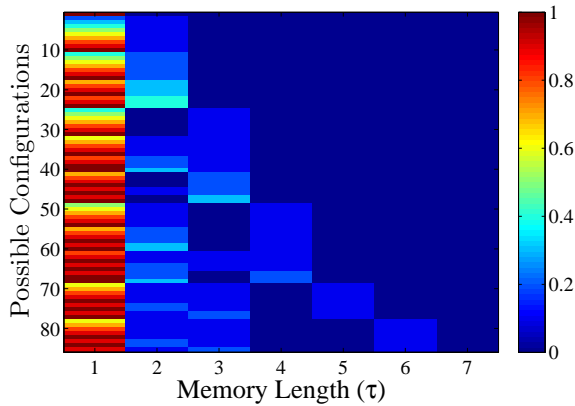}
\caption{\label{fig:1}
(Color online) The configuration of $\K(\tau)$ in Eq.~\ref{eq:thrmod2} contributed to generate the time evolution. The y-axis illustrates the number of the prior states $\tau$. The x-axis demonstrates all possible weights configurations that rebuild the dynamics of the YCC. The summation over weights are not normalized.}
\end{figure}

\section{Boolean networks}

Boolean networks are discrete dynamical systems which are built up  with  interacting binary elements \cite{Kauffman:2003,Aldana:2003,page1990self,green2007emergence}. Technically, a dynamical directed graph with binary assigned nodes and discrete weighted edges is  called  a Boolean network.
The time evolution of the nodes is governed by  so called Boolean functions. Boolean functions compile the input arrows to a node  according to some Boolean rules and extract  the binary outputs which indicate the value of the nodes at the subsequent time step.

One of the most used classes of Boolean functions in modeling biological networks such as gene regulations and neural signalling is  called the threshold function\cite{Lau:2007,Li:2004},  where
depending on whether the sum over all the inputs is higher or lower than a certain threshold, the output will take two different values.

Mathematically, one considers some  $\sigma(t)$ as the state vector of an $n$-node network at time $t$. The Boolean function updates the state of $j$th node and gives  a  value in the next time step $t+\Delta t$: $\sigma(t+\Delta t)$.

A $k$-input Boolean function $\sigma_i(t+\Delta t)=f_i(\sigma(t))$, on site  $i$,  is called   a {\em general threshold function} if there is
a matrix $W=\{w_{ij}\} \in \mathbb{R}^k$ and a threshold
$\theta \in \mathbb{R}$ such that
\begin{equation} \label{eq:thr}
\sigma_i(t+\Delta t) = H ( \sum_{j=1}^k w_{ij} \sigma_j(t) - \theta)
\end{equation}
for all $\sigma \in \{0,1\}^k$, using the step function
$H: \mathbb{R} \rightarrow\{0,1\}$ with
$H(x) = 1$ if and only if $x > 0$.
In other words, the output of the function depends on the weighted sum of its inputs when compared to a certain threshold value. Here below,  we consider the  case of discrete weights $w_j \in \{-1,0,+1\}$ for all inputs
$j$ and a vanishing threshold $\theta=0$.

The inputs of the function are dependent of the previous values of the nodes which are sending signals to it and the weights of the signals which are being sent.
Therefore, the function is a Markov iteration i.e., one can extract all the information which are required in order to build the next step of the dynamics $strictly$  from the present state. In the following sections, we  discuss  and emphasize why and how the states of the networks in the more previous time steps should take part in producing  upcoming states.


\section{Non-Markovian Boolean Dynamics}

In the time evolution of many dynamical systems, some  trace of history is usually  observed.  For instance, one should commonly admit that the expression level of a gene should not be totally independent of its near history's functioning. Specially,  many vital phenomena in living systems are cyclic. {\it Let us restrict ourselves to endogenous aspects}.  Here, we attempt to impose the effect of memory in the dynamics of Boolean networks by adding some terms  relying on previous states of the network  as inputs of the function.
Technically, the term $ \sigma_j(t)$ is to be replaced by a non-Markovian state, as  in Eq.(~\ref{eq:thr}),

\begin{equation} \label{eq:thrmod0}
\sigma_j(t) \rightarrow   \int  \K(\tau) \sigma_j(t-\tau) d\tau,
\end{equation}
where $ \K(\tau)$ is the kernel of Eq.(\ref{eq:thr}) and $\tau$ indicates how long a "continuous length of the memory" can be taken into account.  If $ \K(\tau) \propto \delta (\tau)$ where $\delta (\tau)$ is Dirac delta function Eq.~\ref{eq:thr} will remain unchanged \cite{hilfer2000applications,safdari2015phenomenon,safdari2015aging,boguna2014simulating}.
For a discrete dynamical procedure Eq.~(\ref{eq:thrmod0}) can be considered as

\begin{equation} \label{eq:thrmod1}
\sigma_j(t) \rightarrow \sum_{\tau} \K(\tau) \sigma_j(t-\tau).
\end{equation}
Therefore Equation(~\ref{eq:thr}) is to be replaced by,

\begin{equation} \label{eq:thrmod2}
\sigma_i(t+\Delta t) = H ( \sum_{j=1}^k w_{ij} (\sum_{\tau} \K(\tau) \sigma_j(t-\tau)) - \theta).
\end{equation}
The share of each prior state $\sigma_i(t-\tau)$ is represented with the summation in Equation~(\ref{eq:thrmod1}). That is, $\K(\tau)$ displays the weight of the $\tau$-th former state to generate the  state at time $t$. The Boolean networks which are updated by this type of Boolean functions are called \textit{non-Markovian Boolean networks}.

\section{Memory in Cell cycle}

\begin{figure}
\centering
\includegraphics[width=0.5\textwidth]{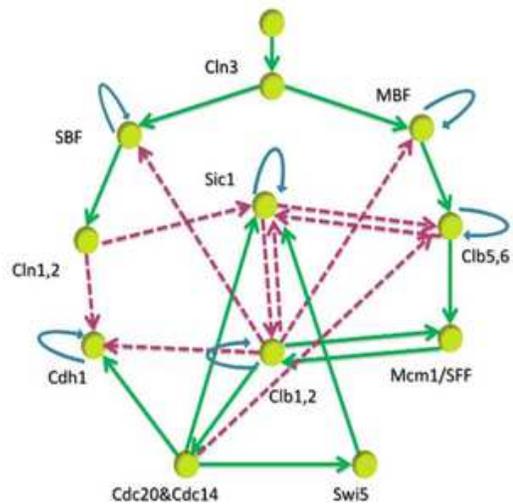}
\caption{\label{fig:bud}
(Color online) Budding YCC network. Solid arrows represent activators,
dashed arrows inhibitors \cite{ebadi:2014}.}
\end{figure}

In order to check the reliability of the idea of memory in empirical Boolean networks, we simulate a cell cycle process first using  a non-Markovian Boolean network. The cell cycle process is a consequence of certain protein-protein interactions
which lead  to a cell division. This vital phenomenon has been  simulated by Boolean networks \cite{nurse:1975,Li:2004,Lau:2007,Davidich:2008,Rybarsch:2012,Boldhaus:2010}. In this system, proteins send chemical signals to activate or deactivate each other. During each step of the cell division process, a certain type of proteins are active and others are inactive. Therefore, each stage of the cell growth process can be addressed as a Boolean state vector. Since the time evolution of the
protein factors that are involved in a cell division process is fairly known, the
state vector sequence of the corresponding network dynamics is  considered to be the most
confirmed information for these systems. The state vector sequence of yeast cell
cycle which terminates to the $G_1$ fixed point is shown in \cite{Li:2004,Lau:2007,Davidich:2008,Rybarsch:2012,Boldhaus:2010}. We study the most remarkable yeast cell cycles (YCC); Budding yeast \emph{Saccharomyces cerevisiae}. The regulatory process that governs the Budding cell cycle is the time evolution of 11 interacting proteins \cite{mendenhall:1998,jorgensen:2004} which can be considered as a protein network (Figure~\ref{fig:bud}).

From a few  simulations, it can be  shown that the modified threshold function (Eq.(\ref{eq:thrmod2})) is capable to successfully generate the dynamical pathway of the Budding yeast cell division based on the given topology. Figure~\ref{fig:1} shows the phase diagram of the weights of all the former states  $\K(\tau)$,  that are involved in generating the cell cycle trajectory of budding YCC network with $13$ time steps length. The $x$-axis shows the number of former states $\tau$ that are contributing in building the dynamics. The $y$-axis shows all possible configurations of the weights of the former states $\K(\tau)$ that are capable to reproduce the trajectory. In other words, we indicate the share of each former state in the dynamics generating. The colors in the diagram illustrate the weights of each former state. The reddish colors correspond to the higher weights while the bluish colors show the lower weights. As the number of former states increases, their weight is reduced so that the weight of the $6$th former state vanishes,  e.g., $\K(6)=0$. Therefore, the maximum number of former states that can contribute in generating the dynamical pathway is equal to $5$, i.e., $\forall \tau_{\in \mathbb{Z}}>5:\K(\tau) =0$.

 On the other hand, as the number of involved former states is reduced, the variety   of  configurations of the rest states also decreases. For instance, there are $85$ different combinations of the former states weights for the situation with $5$ former states are included in Equation~(\ref{eq:thrmod2}) ($\tau=5$). The reddish spectrum in the first column indicates the higher weight of the first former state as compared to the farther posterior states.

The contribution of the former states to generate a part of the cell cycle dynamical pathway (including 7 steps) based on their weights has also  been calculated. The number of configurations is of the order of $\sim10^4$. Apparently, by reducing the length of the dynamical path way which is to be generated, the number of involved former states and configurations of their weights increase. This means  that  the share of former states can be varied over a wide range.

\begin{figure}
\includegraphics[scale=0.5]{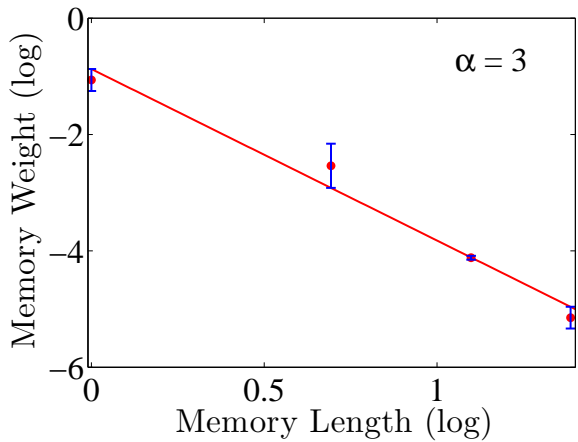}
\caption{\label{fig:4}
The weights configurations with descending order versus the length of memory in a log-log scale. As can be seen from Fig.~\ref{fig:1}, 38 of 85 possible configurations have decreasing memory weight. This diagram is a power law with $\alpha=3$.}
\end{figure}

Now, recall that  complex systems phenomena reveal temporal and/or spatial scalability, - which guarantees their invariant statistical properties through various time and/or length scales \cite{barabasi1999emergence,guo2015memory,goh2008burstiness}. The scalable systems (also called scale free systems) have features characterized by power law distributions. In order to check whether the memory in  usual Boolean dynamics is scale free in  time, we plot the weights in a  descending order (which holds for the memory length up to $\tau=4$) on a log-log diagram (Fig.~\ref{fig:4}). Apparently, 38 of 85 possible configurations have a decreasing memory weight : it can be seen that the weights of the memory states follow a power law with $\alpha=-3$. Therefore, $\K(\tau)$ can be assumed to behave as ,
\begin{equation} \label{eq:thrmod4}
\K(\tau) \propto |\tau|^{-\alpha}.
\end{equation}
Such a  power law behaviour of the weights average of the former states  depicts the scalability of the cell cycle procedure in  time \cite{metzler2000random,Hadiseh,Forough}. Therefore, while the time evolution of cell cycle does not forget the history of its dynamical pathway, its sensitivity to each former state is proportional to a (power of the ) lag time: how far the prior state is from present.

\begin{figure}
\includegraphics[scale=0.5]{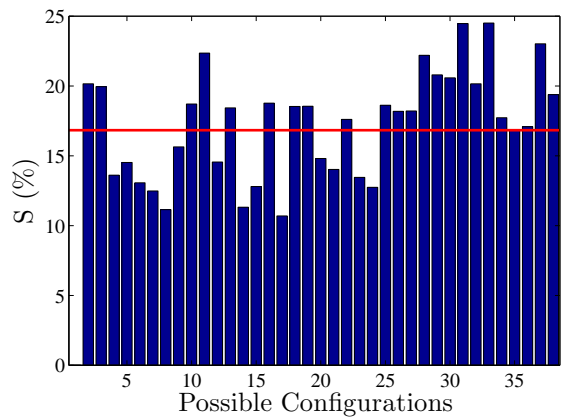}
\caption{\label{fig:6}
The average $S$ over all the elements of the dynamical trajectory for the power law $\K(\tau)$. Y-axis indicates the average $S$ and x-axis illustrates the possible configurations which are shown in Fig.~\ref{fig:1}. The red line is the average of date.}
\end{figure}

\section{Robustness}

A trivial question  can be raised about the ``robustness'' of a  Markovian network dynamics as compared to the non-Markovian ones.
To address this question, in the following,  we define a quantity as a fine measure to compare the robustness of the Markovian dynamics against the non-Markovian one.

Let us assume $\sigma_m$ and $\sigma_{nm}$ indicate the state vectors of the network   updated by a  Markovian function (Equation~(\ref{eq:thr})) and non-Markovian function (Equation~(\ref{eq:thrmod2}))  respectively. In other words, $\sigma_m$ denotes the state of a network with Markovian dynamics while $\sigma_{nm}$ represents the state of a network  in which the prior states play a  role in its time evolution; on the other hand,  $\sigma^{\updownarrow i}_{m}$ denotes the state vector of a Markovian network in which its $i$-th elements is negated. Technically, $\sigma^{\updownarrow i}_{m}$ is defined as 

\begin{equation}\label{eq:negation}
(\sigma^{\updownarrow i})_j \neq \sigma_j \Leftrightarrow i=j ~.
\end{equation}
The difference between the states in which one element is randomly perturbed and an non-perturbed one, can be calculated:  performing the calculation $2^N$ times for both Markovian and non-Markovian systems and taking an average over all of them, leads to a measure which allows to  compare the sensitivity of the two models against perturbations. We define 

\begin{equation} \label{eq:robust}
 S= \frac{(\sigma_{m}-\sigma^{\updownarrow i}_{m})-(\sigma_{nm}-\sigma^{\updownarrow i}_{nm})}{\sigma_m-\sigma^{\updownarrow i}_{m}} \%,
\end{equation}
which serves a criterion for comparison of the robustness of the two models. We have performed the analysis based on the above formula on the Boolean network of the yeast cell division. The dynamical pathway of the YCC network was shown in Figure 2 in  Li et al., \cite{Li:2004}. We have flipped each of the $(2^N)\times N$ elements within the dynamical pathway for both Markovian and non-Markovian networks and observed the effect of each flipping. We calculate the average of $S$ over all the elements for the $85$ possible configuration of weights (Fig.~\ref{fig:1}) that are capable to reproduce the dynamics of the cell division (which includes 13 time steps over the whole $2^{11}$ steps). The percentage  of the difference between the perturbed and non-perturbed dynamics for Markovian and non-Markovian networks is positive for  most of the configurations:  $(\sigma_{m}-\sigma^{\updownarrow i}_{m})>(\sigma_{nm}-\sigma^{\updownarrow i}_{nm})$. This means that the Markovian dynamics are more sensitive to small perturbation than non-Markovian dynamics. 
 
 The average of $S$ over all configurations with different memory length is $\simeq 13 \%$ (for 85 possible configurations). This observation indicates the higher robustness of non-Markovian dynamics compare to the Markovian one. Therefore, memory appears as a positive factor in raising the stability of this dynamical process.

Similarly, the average of $S$ over all elements for all possible configurations of $\K(\tau)$ that follow the power law distribution is shown in Figure~\ref{fig:6} where the average is $\simeq 17 \%$ (for 38 possible configurations). However the average over the weight configurations which do not follow a descending pattern is $\simeq 4 \%$ (for 47 possible configurations). Thus, while the power law memory significantly increases the dynamical robustness of the cell cycle Boolean network, the randomly distributed memory does not considerably influence the dynamical robustness of the system.
Thereafter, it can be conjectured that a ``non power law memory'' might be considered as some noise which does not play an effective role in the dynamical process of non-Markovian Boolean networks.

\section{Discussion and conclusion}

In this article, we have studied non-Markovian Boolean dynamics in order  to investigate the influence of memory in the dynamical path of a Boolean network.
First , we have shown the capability of the non-Markovian threshold function to simulate the cell cycle regulatory network. Although inserting the memory terms in Boolean function causes a loss in Boolean discretizing simplification approach, it may lead to a more realistic model for simulating  biological process. In the next step, we have investigated the effect of perturbations in Markovian and non-Markovian Boolean networks. We conclude that non-Markovian Boolean dynamics reveals a much more robust behaviour as compared to the Markovian ones. As a significant achievement, we extract a power law memory in the dynamics of non-Markovian cell cycle network. This observation is consistent with the nature of self organizing properties of biological systems. Thus, while the system carries its time evolution information through time, a perturbation in the farther prior states cannot deviate the system from its dynamical pathway.

The idea of non-Markovian Boolean network can be further explored, in particular in molecular biology self organizing systems as those considered in\cite{kauffman1996home,camazine2003self}.



\end{document}